Publication date: December 2021
Embargo: 24 Months
European Union, Horizon 2020, Grant Agreement number: 857470 — NOMATEN — H2020-WIDESPREAD-2018-2020
DOI: https://doi.org/10.1016/j.ceramint.2021.09.013

# Absolute radiation tolerance of amorphous alumina coatings at room temperature

A. Zaborowska[a,*], Ł. Kurpaska[a], M. Clozel[a,+], E.J. Olivier[b], J.H. O'Connell[b], M. Vanazzi[c,d], F. Di Fonzo[c], A. Azarov[a,e], I. Jóźwik[a,f], M. Frelek-Kozak[a], R. Diduszko[f], J.H. Neethling[b], J. Jagielski[a,f]

[a] *National Center for Nuclear Research, A. Sołtana 7, 05-400 Otwock-Świerk, Poland*
[b] *Centre for High Resolution Transmission Electron Microscopy, Physics Department, Nelson Mandela University, PO Box 77000, Port Elizabeth, 6031, South Africa*
[c] *Center for Nano Science and Technology @PoliMi, Istituto Italiano di Tecnologia, Via Giovanni Pascoli 70/3, 20133 Milano, Italy*
[d] *Dipartimento di Energia, Politecnico di Milano, Via Ponzio 34/3, 20133, Milano, Italy*
[e] *Department of Physics, Centre for Material Science and Nanotechnology, University of Oslo, P.O. Box 1048, Blindern, N-0316 Oslo, Norway*
[f] *Institute of Electronic Materials Technology, st. Wólczynska 133, 01-919 Warsaw, Poland*
[+] *Present address: Institut für Materialphysik im Weltraum, Cologne, Germany*

*corresponding author: agata.zaborowska@ncbj.gov.pl, Phone: (+48) 22 273 1061, Fax: (+48) 22 273 1061*

Abstract

In this study structural and mechanical properties of a 1 µm thick $Al_2O_3$ coating, deposited on 316L stainless steel by Pulsed Laser Deposition (PLD), subjected to high energy ion irradiation were assessed. Mechanical properties of pristine and ion-modified specimens were investigated using the nanoindentation technique. A comprehensive characterization combining Transmission Electron Microscopy and Grazing-Incidence X-ray Diffraction provided deep insight into the structure of the tested material at the nano- and micro-scale. Variation in the local atomic ordering of the irradiated zone at different doses was investigated using a reduced distribution function analysis obtained from electron diffraction data. Findings from nanoindentation measurements revealed a slight reduction in hardness of all irradiated layers. At the same time TEM examination indicated that the irradiated layer remained amorphous over the whole dpa range. No evidence of

**Accepted Version**

Publication date: December 2021
Embargo: 24 Months
European Union, Horizon 2020, Grant Agreement number: 857470 — NOMATEN — H2020-WIDESPREAD-2018-2020
DOI: https://doi.org/10.1016/j.ceramint.2021.09.013

crystallization, void formation or element segregation was observed up to the highest implanted dose. Reported mechanical and structural findings were critically compared with each other pointing to the conclusion that under given irradiation conditions, over the whole range of doses used, alumina coatings exhibit excellent radiation resistance. Obtained data strongly suggest that investigated material may be considered as a promising candidate for next-generation nuclear reactors, especially LFR-type, where high corrosion protection is one of the highest prerogatives to be met.

1. Introduction

Successful development of materials for next generation nuclear systems continues to be a major challenge. Among the six approved by the GIF (Generation IV International Forum) reactor concepts, LFR (Lead-Cooled Fast Reactor) seems to be among prime candidates to be built in the next decade. This particular solution offers also great improvements over existing nuclear energy systems in terms of sustainability and economics [1,2]. It is characterized by an advanced fuel cycle and an increase of fuel burnup levels will aid greatly towards sustainability goals. However, in order to reach high fuel burnup, it is necessary to maximize the time during which the fuel remains in the reactor core. In this way the amount of generated specific energy can be raised, and the long-term radiotoxicity of irradiated nuclear fuel be significantly reduced.

**Accepted Version**

Publication date: December 2021
Embargo: 24 Months
European Union, Horizon 2020, Grant Agreement number: 857470 — NOMATEN — H2020-WIDESPREAD-2018-2020
DOI: https://doi.org/10.1016/j.ceramint.2021.09.013

In order to deploy this technology and construct first operational units one must select materials, from which these systems will be built. During the last decade, a number of investigations have been carried out regarding corrosion resistance, mechanical and structural properties or radiation resistance [2–8]. Among many critical parts of the LFR type reactor, special attention should be given to understand the performance of the fuel cladding. For example, it is anticipated that these parts will receive in relatively short amount of time very high irradiation fluencies (~100-200 dpa), will have to retain its structural integrity at high temperature and operate under highly corrosive media attack [9]. Unfortunately, there is currently a lack of suitable materials able to withstand the expected harsh operating conditions (high levels of neutron fluence, high temperatures and contact with liquid lead) [7].

Taking into account all these requirements, ceramics represent a very promising class of materials due to their high temperature strength and excellent corrosion resistance combined with their structural integrity up to large dpa's [10]. However, the main issue with ceramics is their limited plasticity at low temperatures [11]. In addition to that, this behaviour must be well documented and understood for irradiated material. One may consider necesity to cool down the reactor quickly. Therefore, no delamination, cracking or peeling from the metallic substrate should take place.

Such studies have been started recently and they suggest that for Lead-cooled Fast Reactor (LFR) systems the use of alumina nanocrystalline coatings can offer a potential solution to these issues [10–12]. However, the implementation of new materials solutions requires a series of systematic tests (preferably at simulated irradiation conditions) to

**Accepted Version**
Publication date: December 2021
Embargo: 24 Months
European Union, Horizon 2020, Grant Agreement number: 857470 — NOMATEN — H2020-WIDESPREAD-2018-2020
DOI: https://doi.org/10.1016/j.ceramint.2021.09.013

assess the material performance. The purpose of these tests is to ensure that the material will retain its structural integrity within specified tolerances throughout its service lifetime. The traditional approach consisting of material irradiation in a test-reactor core and subsequent post-irradiation characterization using standard methods can be time-consuming and costly. This is due to the limited amount of neutron sources and radioactivity of an investigated sample [13]. An alternative procedure involves the use of ion irradiation (as a surrogate for neutron irradiation) followed by small-scale testing [14–18]. Partially, that was performed in previous work [19] where the influence of room temperature (RT) ion irradiation on the nanomechanical properties of thin alumina coatings deposited by PLD (Pulsed Laser Deposition) was investigated. Broadly speaking, nanomechanical investigations revealed a minor decrease of the hardness of the studied coatings upon ion irradiation. It should be noted that the knowledge about the behavior of the alumina under room temperature irradiation is of vital importance for complete understanding of its behavior under irradiation. This is due to the fact that the phenomena occurring in the material are considered to be the result of the synergistic effect of high and room temperature irradiations [11]. Therefore, the only way to understand a whole picture of the in-service behavior of alumina is to investigate its properties in different irradiation and temperature conditions, and study them independently.

This paper aims to corroborate the findings presented in [19]. Previous work was mainly focused on nanomechanical investigations and was limited to low-energy ion irradiation (250 keV). For this study, high energy (1.2 MeV) ion beam implantation was

**Accepted Version**
Publication date: December 2021
Embargo: 24 Months
European Union, Horizon 2020, Grant Agreement number: 857470 — NOMATEN — H2020-WIDESPREAD-2018-2020
DOI: https://doi.org/10.1016/j.ceramint.2021.09.013

performed, resulting in more clarity of the nanoindentation data obtained. Furthermore, the research was expanded by conducting a comprehensive structural analysis of irradiated layers using TEM and GIXRD. Additionally, for the first time the local atomic-ordering of irradiated layers were investigated by performing a reduced distribution function [20] analysis on electron diffraction data obtained from irradiated zones. This approach provided a deeper understanding of the structural evolution of alumina under applied high energy irradiation. We also compared obtained mechanical and structural data, which provided new insight into the understanding of the amorphous alumina coating radiation resistance.

2. Experimental

   2.1. Materials preparation

Thin film of alumina coating was grown on 316L stainless steel (SS) substrates using the PLD (Pulsed Laser Deposition) technique. Steel substrates (1 cm$^2$ plates) were cut by WEDM (Wire Electrical Discharge Machining) from an annealed sheet with a thickness of 0.9 mm. It is known that PLD-grown coatings generally reproduce the roughness of the substrate surface [21], which for certain measurement techniques (nanoindentation) may constitute a challenge in delivering credible data. For that reason, to obtain a smooth and suitable coating for further investigation, the steel samples were mirror-polished before the deposition process started. Polishing included standard metallographic steps, i.e. the use of SiC abrasive papers (up to 1200 grade) followed by using a colloidal silica suspension (0.06 μm grain size). Detailed information regarding the deposition process can be found elsewhere [21]. The thickness of the investigated

**Accepted Version**
Publication date: December 2021
Embargo: 24 Months
European Union, Horizon 2020, Grant Agreement number: 857470 — NOMATEN — H2020-WIDESPREAD-2018-2020
DOI: https://doi.org/10.1016/j.ceramint.2021.09.013

coating was approx. 1 μm. However, for the methodological purposes, one extra specimen with a 5 μm thick coating was prepared and tested.

2.2. Ion irradiation and SRIM calculations

As-manufactured alumina-coated samples were ion-irradiated at room temperature with 1.2 MeV $Au^+$ ions at six fluences varying from approx. $8.0 \times 10^{13}$ cm$^{-2}$ to $4.0 \times 10^{15}$ cm$^{-2}$. Fluences corresponded to damaging levels of 0.5; 1; 3; 5; 10 and 25 dpa (displacement per atom [22]) were calculated by using SRIM (The Stopping and Range of Ions in Matter) software [23]. When designing irradiation conditions, importance was placed on limiting the contribution of electronic energy loss. This is considered to be a key factor for reliable simulations of neutron-induced damage in ceramic insulators [24–26]. In order to properly express the amount of ionization, the electronic-to-nuclear stopping power (ENSP) ratio was used. Generally speaking, lower ENSP ratios are obtained for heavier ions, contrary to lighter ions of the same energy [24]. Therefore, in order to properly simulate the impact of neutron damage in alumina coating, gold ions with a high atomic mass were selected. The calculated ratio of ENSP is below 5, while the maximum ionizing component of the irradiation spectrum in this case is below 3.7 keV/nm. It should be noted that SRIM predictions for heavy ions in lighter targets are affected by large errors arising from an overestimation of the electronic stopping power [27]. Experimental studies show that ion ranges obtained after heavy ions implantation with MeV initial energy into light targets can be higher than TRIM predicted values by 20 to even 50% [27]. Therefore, all of the above mentioned values could be burdened with some dose of an unavoidable uncertainty. As can be seen from Fig. 1, for

Publication date: December 2021
Embargo: 24 Months
European Union, Horizon 2020, Grant Agreement number: 857470 — NOMATEN — H2020-WIDESPREAD-2018-2020
DOI: https://doi.org/10.1016/j.ceramint.2021.09.013

the given irradiation conditions, the radiation and projected ranges in the investigated coating are approx. 270 nm and 300 nm, respectively. The ion implantation campaign was carried out at the Norwegian Micro- and Nano- Fabrication Facility, NorFab in 9SDH-2 NEC Pelletron Tandem accelerator.

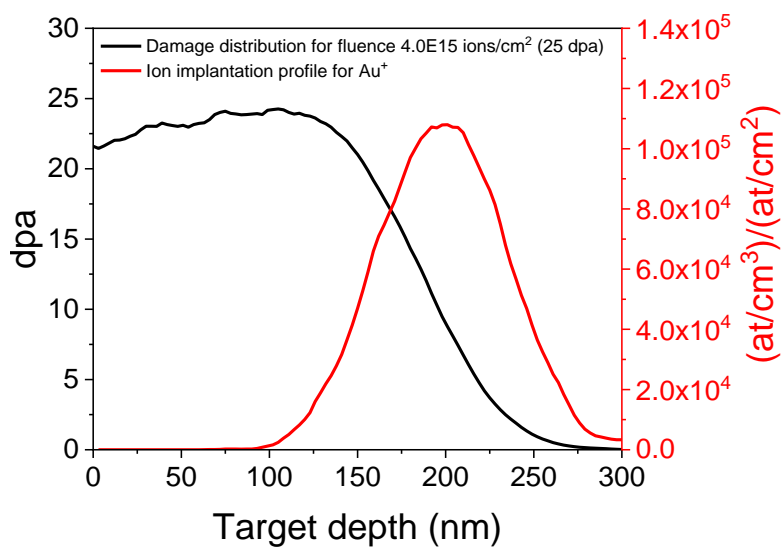

Fig. 1. Damage and ion implantation profiles of alumina coating submitted to irradiation up to $4.0 \times 10^{15}$ ions/cm$^{-2}$.

2.3. Scanning Electron Microscopy (SEM)

The scanning electron microscopy observations were carried out using a Hitachi SU8230 and Carl Zeiss Auriga CrossBeam Workstation, both equipped with advanced in-lens detection systems. By using the high-resolution low-kV imaging, the topographical contrast and the surface morphology of the virgin and 25 dpa ion irradiated samples were revealed. Studied specimens were sputter-coated with a sub-nanometer-thick conductive layer to avoid charging during observations. In order to avoid

**Accepted Version**
Publication date: December 2021
Embargo: 24 Months
European Union, Horizon 2020, Grant Agreement number: 857470 — NOMATEN — H2020-WIDESPREAD-2018-2020
DOI: https://doi.org/10.1016/j.ceramint.2021.09.013

uncertainty, structural investigations and conductive layer deposition were performed after nanoindentation tests.

2.4. Nanoindentation

In order to investigate the influence of ion irradiation on the nanomechanical properties of the investigated coatings, series of nanoindentation tests were performed using NanoTest Vantage system (Micro Materials Ltd). All tests were performed at room temperature using a Berkovich-shaped diamond indenter. To minimize the errors arising from rounding (and other deviations) from the ideal shape of the indenter tip, the actual diamond area function (DAF) of the indenter tip at different indentation depths was determined before testing commences. Nanohardness (H) and reduced Young's modulus ($E_r$) were extracted from the unloading data, using well known Oliver and Pharr model [28]. To convert reduced Young's modulus to Young's modulus, a Poisson ratio of 0.29 [21] was used.

The volume of the plastically deformed zone surrounding the indentation tip is an important parameter for nanoindentation data interpretation, particularly when testing thin films [16]. The knowledge about plastic zone size allows for the design of nanoindentation test parameters so that the contribution from substrate deformation is negligible. Admittedly, the commonly accepted rule states that the plastic zone associated with nanoindentation extends to a maximum of about 10 times of the plastic penetration depth. However, this is a somewhat simplified assumption, since this factor can range from 2 to 10 and is dependent on several aspects such as the material type, crystal orientation, tip geometry, indentation depth, surface residual stress and yield strength of

**Accepted Version**
Publication date: December 2021
Embargo: 24 Months
European Union, Horizon 2020, Grant Agreement number: 857470 — NOMATEN — H2020-WIDESPREAD-2018-2020
DOI: https://doi.org/10.1016/j.ceramint.2021.09.013

the material [16]. Rather than following the above-mentioned rule, an alternative approach can be applied. One possibility is to perform preliminary indentations at increasing penetration depths. This approach gives the relation between the plastic penetration depth and size of the plastic zone which extends beneath the indenter tip [17,29].

For this purpose, as the first step, the 5 μm virgin coating was tested by using Multiple Load Cycle mode with increasing load. The use of the thicker coating was intentional and aimed to avoid the errors connected with shallow indentations, thus ensuring the clarity and unambiguous data interpretation (which is particularly important for thinner, ion irradiated specimens). A single indentation included 20 cycles, with increasing force from 5 to 500 mN. The following indentation cycle sequence was implemented: loading with the rate of 2.5 mN/s, holding at maximum load for 10 s, and unloading with the rate of 2.5 mN/s to 75 % of the maximum load of the ongoing cycle. Measurement of a thermal drift at the end of each indentation, before complete unloading was also performed. In this study, the most crucial is the behavior of the material at lower forces. Therefore, maximum loads were selected so that they are more numerous at shallower depths, i.e. 5; 25; 40; 55; 70; 85; 100; 115; 130; 145; 160; 175; 190; 205; 225; 245; 270; and rare for the highest loads i.e. 300; 400; 500 mN. A series of 10 indentations were performed in each case. Finally, the calculated hardness values were plotted against depth.

In the second step, the actual subject of the investigation (1 μm coating) was tested under analogous conditions (i.e. on Multiple Load Cycle mode with increasing load).

**Accepted Version**
Publication date: December 2021
Embargo: 24 Months
European Union, Horizon 2020, Grant Agreement number: 857470 — NOMATEN — H2020-WIDESPREAD-2018-2020
DOI: https://doi.org/10.1016/j.ceramint.2021.09.013

Single indentation included 16 cycles with the force increasing from 0.2 to maximum 20 mN. Here, the following indentation cycle sequence was implemented: loading with the rate of 0.5 mN/s, holding at maximum load for 2 s, and unloading with the rate of 0.5 mN/s to 75% of the maximum load of the ongoing cycle. Each indent also had a thermal drift measurement at the end of the indentation, before complete unloading. The maximum loads included: 0.2; 0.3; 0.4; 0.5; 0.7; 1; 1.2; 1.5; 2; 2.5; 3; 5; 7; 10; 15; 20 mN. Such measurements were performed for virgin and ion-irradiated samples (up to 25 dpa).

In the final step, all of the six irradiated specimens and pristine material were tested using load-controlled mode, under a single force, determined from the given irradiated layer depth and the findings of the two preceding steps. Each indentation was repeated 20 times, with an interval of 50 μm between the indents. One must remember that each data point reported in the Fig. 3-4 and 6 represents the average value for at least 10 measurements while error bars represent the calculated standard deviation.

### 2.5. Structure characterization

Comprehensive structural characterization of the PLD-grown alumina coating before and after ion-irradiation was obtained by using a combination of Transmission electron microscopy (TEM) and Grazing Incidence X-Ray Diffraction (GIXRD) techniques. These methods are complementary and they provided us a detailed information about the material structure which resulted in an understanding of the material structural integrity. Finally, this approach also provided insight into the structural evolution of the alumina upon irradiation.

#### 2.5.1. Transmission electron microscopy (TEM)

**Accepted Version**

Publication date: December 2021
Embargo: 24 Months
European Union, Horizon 2020, Grant Agreement number: 857470 — NOMATEN — H2020-WIDESPREAD-2018-2020
DOI: https://doi.org/10.1016/j.ceramint.2021.09.013

Electron transparent lamellae were prepared by a standard FIB lift-out technique using an FEI Helios Nanolab 650. Final thinning of the lamellae was performed with 5 keV Ga ions followed by a 2 keV Ga polish. Electron transparent lamellae were analyzed in a double Cs corrected JEOL ARM 200F equipped with an Oxford Xmax100 EDS detector and Gatan GIF 965ERS with dual EELS capability, operating at 200 kV. The analysis was performed in both TEM and STEM mode. Imaging and analysis were done in scanning transmission (STEM) mode using a sub-angstrom sized probe with a probe current between 68 pA and 281 pA. Probe current conditions were selected in such a way to optimize the beam current but, at the same time, minimize the risk of electron beam damage to the specimen. The convergence semi-angle of the probe was fixed at 23 mrad. The BF detector acceptance semi-angle was set at 0 to 12 mrad by using an illumination limiting aperture. The selected area electron diffraction (SAED) analysis was performed in TEM mode using parallel illumination, a 50 μm condenser aperture and selected area aperture including a circular area with radius approximately 100 nm. Extraction of the reduced distribution functions from diffraction patterns was done with the commercial software eRDF Analyser [20]. Fitting of the experimentally determined intensity profiles were done using atomic scattering factors as described in Lobato et al. [30] and inter-atomic distances (r) of 0.7 nm and less. Further information related to the fitting routine used by the software may be found in the paper by Shanmugam et al. [20].

2.5.2. Grazing Incidence X-Ray Diffraction (GIXRD)

X-ray diffraction was undertaken using a Rigaku SmartLab X-ray diffractometer with Cu-Kα radiation (λ≈1.5418 Å) produced at 40 kV and 30 mA. The data were

**Accepted Version**
Publication date: December 2021
Embargo: 24 Months
European Union, Horizon 2020, Grant Agreement number: 857470 — NOMATEN — H2020-WIDESPREAD-2018-2020
DOI: https://doi.org/10.1016/j.ceramint.2021.09.013

collected at room temperature in a parallel beam (PB) mode in the range from 20º to 80º of 2θ angle and with an incidence angle α of 0.5º. The software program to analyze the data was Rigaku PDXL program with the ICDD database PDF4 + 2018. The effective penetration depth τ was calculated based on the equation [31]:

$$\tau = \frac{-ln(1 - G_x)}{\mu(\frac{1}{sin\alpha} + \frac{1}{\sin(2\theta - \alpha)})}$$

where $G_x$ corresponds to the absorbed fraction of the total intensity of the X-ray beam. Here $G_x$ is defined as $G_x$=1-1/10=0.9. For the calculations, the linear attenuation coefficient $\mu$ = 97 cm$^{-1}$ [32] was taken. It was estimated that for the wavelength of 1.54 Å the penetration depth in alumina is approx. 2 μm. Thus, keeping in mind the depth of ion-irradiated layer (approx. 225 nm) and the thickness of the coating (approx. 1 μm), one must keep in mind that portion of the detected signal originates from the steel substrate and the unmodified coating.

3. Results

3.1. SEM (Scanning Electron Microscopy)

The SEM images in topography contrast of the PLD-grown alumina coating before and after 25 dpa irradiation are presented in Fig. 2. The images show that the surfaces of both samples are smooth and free of large discontinuities. The surface impurities/contaminants or imperfections present at the sample surface are visible as darker areas in the collected images and are not relevant for any quantitative analysis. The comparison of the images collected on both, irradiated and non-irradiated samples

**Accepted Version**
Publication date: December 2021
Embargo: 24 Months
European Union, Horizon 2020, Grant Agreement number: 857470 — NOMATEN — H2020-WIDESPREAD-2018-2020
DOI: https://doi.org/10.1016/j.ceramint.2021.09.013

provide evidence that ion irradiation does not induce any visible changes into the coating surface morphology.

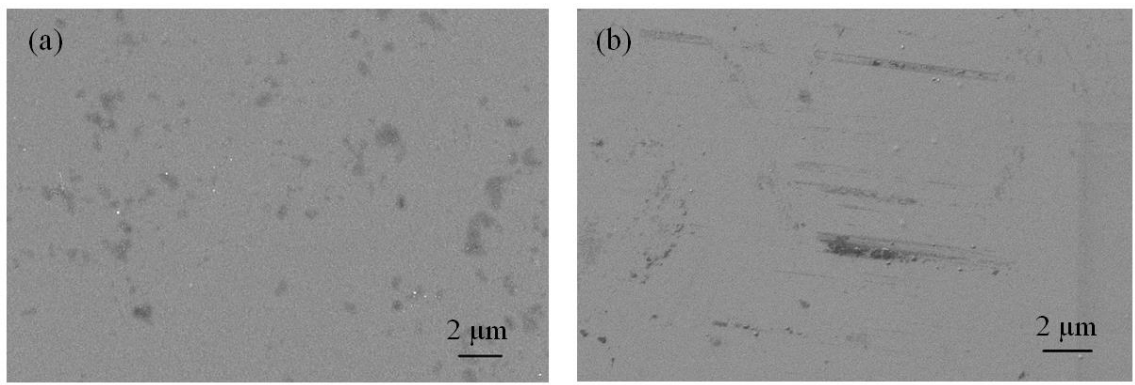

Fig. 2. SEM images of PLD-grown alumina coating surface: (a) in the pristine state and (b) after 25 dpa ion irradiation (1.2 MeV $Au^+$).

3.2. Nanoindentation

In order to determine the impact of ion irradiation on the mechanical properties of the coating, the nanoindentation technique was implemented. The target measurements were preceded by preliminary tests aimed at determining the indicative relation between the plastic penetration depth and size of the plastic zone extending beneath the indenter tip in alumina. Fig. 3 shows the nanohardness of the virgin 5 μm alumina coating plotted as a function of the indentation depth. As can be seen in Fig. 3, the following regions can be distinguished in the curve: initial rise (I), plateau (II) and fall-off (III). The presence of all three regimes confirms the validity of the obtained data and indicates that the range of applied loads was selected suitably [29]. The initial rise (I) in hardness is caused by the fact that for low penetration depths the indenter tip pressed into the material is initially round and so most of the deformation may originate from the elastic portion of the

**Accepted Version**

Publication date: December 2021
Embargo: 24 Months
European Union, Horizon 2020, Grant Agreement number: 857470 — NOMATEN — H2020-WIDESPREAD-2018-2020
DOI: https://doi.org/10.1016/j.ceramint.2021.09.013

material response. Therefore, one must remember that the mean contact pressure, which is the measure of hardness, does not reflect the actual film hardness under such conditions. This phenomenon is well described in the literature [29] and is particularly pronounced for very hard materials. In the next step, measurements were taken at deeper depths, when the plastic zone is fully developed. In this case hardness reaches a plateau tendency (II). The slight hardness decrease observed within this range can be assigned to the well-known indentation size effect (ISE) [33–36], which exists also in amorphous materials [37]. Finally, for the deepest indents, since the plastic zone spreads into the softer steel substrate, a sharp hardness decent (III) is recorded. This effect is known as the softer substrate effect (SSE) [33,36]. In conclusion, the most important information to be extracted from this graph is that for this system, the multiplication factor relating the plastic nanoindentation depth with the plastic zone size is about 7.5. This is evidenced by the fact that the hardness decrease resulting from SSE starts to be pronounced for the plastic depths corresponding to approx. 13 % of the coating thickness (i.e. the line between region II and III, in Fig. 3).

**Accepted Version**

Publication date: December 2021
Embargo: 24 Months
European Union, Horizon 2020, Grant Agreement number: 857470 — NOMATEN — H2020-WIDESPREAD-2018-2020
DOI: https://doi.org/10.1016/j.ceramint.2021.09.013

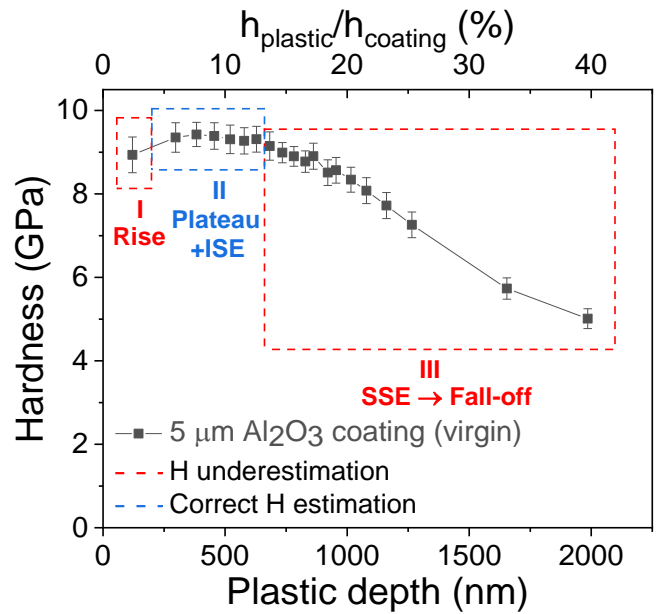

Fig. 3. Nanohardness of virgin 5 μm alumina coating (on 316L SS) versus plastic indentation depth. Dashed boxes marked on the graph indicate the following regions: initial rise (I), plateau (II) and fall-off (III). Measurements made in multi-force mode.

Fig. 4 shows the depth profile for 1 μm alumina coating before (black curve) and after $Au^+$ irradiation with 1.2 MeV (red curve). One can see that the presented curves are composed of parts analogous to those discussed in the preceding subparagraph. It should also be noted that the hardness of the irradiated coating is lower than in virgin state across the entire range of applied forces.

In order to perform detailed analysis of ion irradiation impact on the nanomechanical properties of alumina coating, the force of 1 mN was selected. Given that this load corresponds to the indentation depth of about 42 nm, which represents approx. 16 % of ion irradiated layer we assume that most of the recorded signal originates from damaged zone. However, one should be aware that some (although very minor)

**Accepted Version**
Publication date: December 2021
Embargo: 24 Months
European Union, Horizon 2020, Grant Agreement number: 857470 — NOMATEN — H2020-WIDESPREAD-2018-2020
DOI: https://doi.org/10.1016/j.ceramint.2021.09.013

impact of the tip-rounding and the unmodified bulk alumina coating can be detected. Nevertheless, the adopted methodology seems to be correct and the limits impact of incomplete contact of the indenter tip to the material surface – which for such analysis is of primordial importance. In addition to that, adopted measurement methodology confirms that for the selected testing parameters, no influence of the steel substrate on the hardness values was obtained.

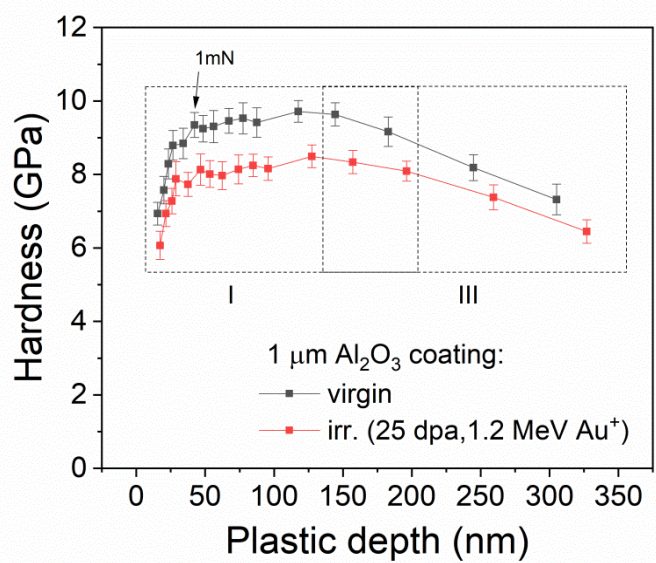

Fig. 4. Nanohardness versus plastic indentation depth for 1 µm alumina coating (on 316L SS): in virgin state (black curve), and after ion irradiation with 1.2 MeV Au$^+$ up to 25 dpa (red curve). Dashed boxes marked on the graph indicate the following regions: initial rise (I), plateau (II) and fall-off (III). Measurements made in the multi-force mode.

Fig. 5 presents a schematic illustration of the measurements conducted on 1 µm ion irradiated coatings under the test conditions indicated above. It is apparent from Fig. 5 that by taking advantage of high energy implantation, the obtained nanomechanical

**Accepted Version**
Publication date: December 2021
Embargo: 24 Months
European Union, Horizon 2020, Grant Agreement number: 857470 — NOMATEN — H2020-WIDESPREAD-2018-2020
DOI: https://doi.org/10.1016/j.ceramint.2021.09.013

results of ion irradiation material are robust and reliable. Therefore, reported herein mechanical data can be treated as representative information about the impact of simulated neutron damage.

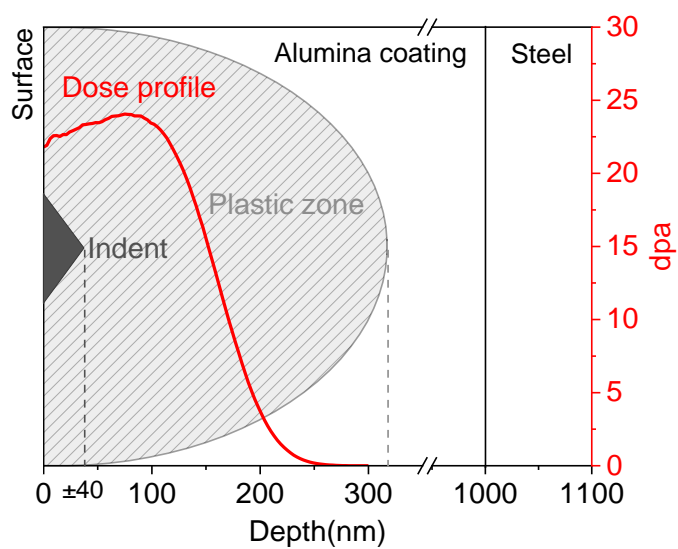

Fig. 5. A schematic illustration of nanoindentation test of 1 μm thick ion-irradiated coating. It is assumed that the depth of the plastic zone is approx. 7.5x of the nanoindentation depth and mainly covers the irradiated region.

Fig. 6a presents the nanohardness and Young's modulus of alumina coating as a function of radiation damage. The representative L-D curves of nanoindentation measurements are shown in Fig. 6b. The as-deposited coating is characterized by nanohardness of H = 10.2 ± 0.3 GPa, which is consistent with previous results reported by G. Ferré et al. [21]. As can be seen in Fig. 6, after room temperature ion irradiation, reduction in hardness has been observed. The relationship between hardness and irradiation dose is not monotonic. Reported descent varies depending on the damage

**Accepted Version**
Publication date: December 2021
Embargo: 24 Months
European Union, Horizon 2020, Grant Agreement number: 857470 — NOMATEN — H2020-WIDESPREAD-2018-2020
DOI: https://doi.org/10.1016/j.ceramint.2021.09.013

level: the lowest decrease occurs around 0.5 dpa (H = 9.6 ± 0.3 GPa, 6%) while the highest value is recorded around 3 dpa (H = 9.1 ± 0.2 GPa, 11 %). One should bear in mind that since the damage profile (Fig. 1) is not homogenous, it is not possible to assign a specific hardness value to a single dose [35]. When a nanoindentation measurement of an ion-irradiated sample is performed, a wide range of doses is sampled. Previous studies showed that the formation of a crystalline phase is accompanied by a hardness increase [10,11,24], thus it is expected that the structural analysis will not show any evidence of crystallization. As for Young's modulus, from Fig. 6a it is apparent that changes due to ion irradiation are negligible and are within error ranges. Pristine coating is characterized by constant Young's modulus of E = 194 ± 9 GPa, which is in a very good agreement with the previously reported values [19,21], when measurements were made using both nanoindentation and Brillouin spectroscopy techniques. Experimental results suggest that Young's modulus value remains constant up to large radiation damage when room temperature ion irradiation is performed.

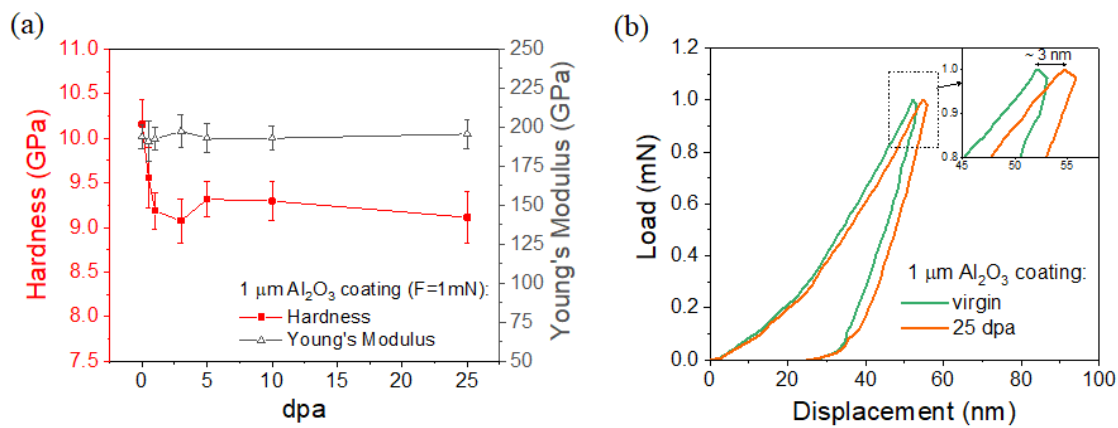

Fig. 6. (a) Nanohardness and Young's modulus of 1 μm alumina coating (on 316L SS) versus irradiation dose. (b) Representative load-displacement curves measured on the

**Accepted Version**
Publication date: December 2021
Embargo: 24 Months
European Union, Horizon 2020, Grant Agreement number: 857470 — NOMATEN — H2020-WIDESPREAD-2018-2020
DOI: https://doi.org/10.1016/j.ceramint.2021.09.013

pristine and irradiated with 1.2 MeV Au$^+$ samples. Measurements were made in constant load (1 mN) mode.

### 3.3. Transmission electron microscopy (TEM)

In order to examine the impact of ion irradiation on the atomic structure of the alumina layer, the TEM technique was used. Three alumina-coated samples were selected for this analysis: virgin, 3 dpa (exhibiting the highest change in hardness) and 25 dpa (irradiated to the highest damage level). All TEM measurements were made on the cross-sectioned specimens. Fig. 7 shows a bright-field (BF) TEM micrograph of the virgin alumina coating. Fig. 8 and 9 show high angle annular dark-field (HAADF) scanning transmission electron microscopy (STEM) images of the alumina coating after 3 and 25 dpa irradiations, respectively. Both profiles consist of two regions: (I) irradiated layer (denoted by the red arrow) and (II) unmodified bulk, material with an as-manufactured coating thickness of approximately 1.1 μm.

A selected area electron diffraction (SAED) pattern was collected from every tested specimen. Measurements were performed in ion irradiated and unmodified regions. Fig. 7 shows signal registered in the virgin alumina coating which indicates that the specimen has amorphous structure. This is confirmed by the presence of diffuse rings in the diffraction pattern. SAED patterns (Figs. 8 and 9) obtained from the irradiated regions (3 and 25 dpa) showed no signs of crystallization with SAED patterns, once again proving that the material is characterized by amorphous structure.

Quantitative EELS elemental mapping (not shown) of the irradiated layers showed no signs of elemental segregation within the detection limit of the technique (~ 1 at%) with

**Accepted Version**
Publication date: December 2021
Embargo: 24 Months
European Union, Horizon 2020, Grant Agreement number: 857470 — NOMATEN — H2020-WIDESPREAD-2018-2020
DOI: https://doi.org/10.1016/j.ceramint.2021.09.013

images (BF, TEM, HAADF) of the layers showing no contrast consistent with the presence of irradiation-induced voids. In Fig. 10, high-resolution TEM (HRTEM) images for virgin and 25 dpa irradiated alumina-coated samples are compared. No evidence for the presence of atomic ordering was observed. Fig. 10 shows high-resolution TEM (HRTEM) images of virgin and 25 dpa irradiated alumina samples. Once again, no evidence of atomic ordering was observed. Obtained results clearly indicate that the coating retain its amorphous structure and structural integrity, despite being submitted to relatively large radiation damage (25 dpa).

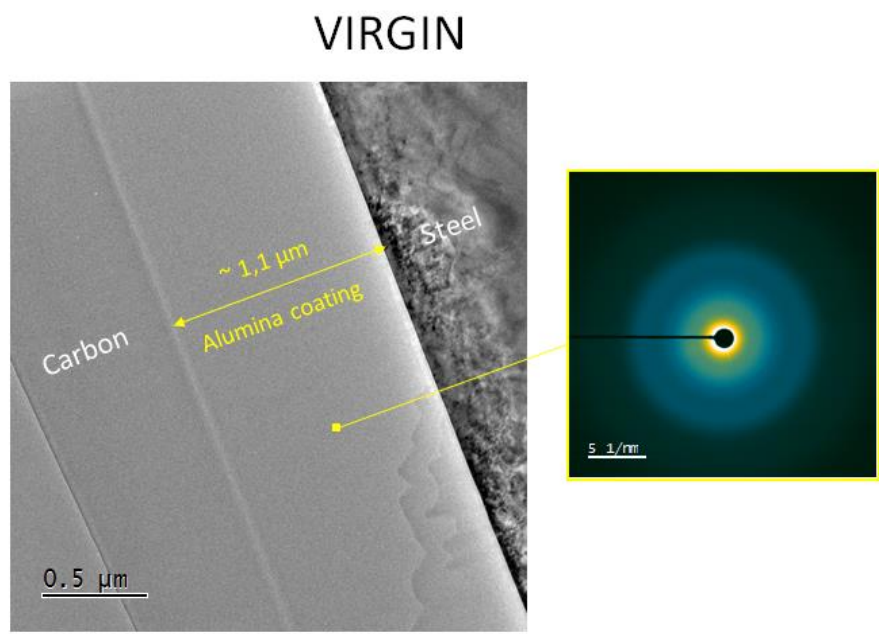

Fig. 7. BF TEM cross-sectional micrograph showing the alumina coating in the pristine state and b) corresponding DP.

**Accepted Version**
Publication date: December 2021
Embargo: 24 Months
European Union, Horizon 2020, Grant Agreement number: 857470 — NOMATEN — H2020-WIDESPREAD-2018-2020
DOI: https://doi.org/10.1016/j.ceramint.2021.09.013

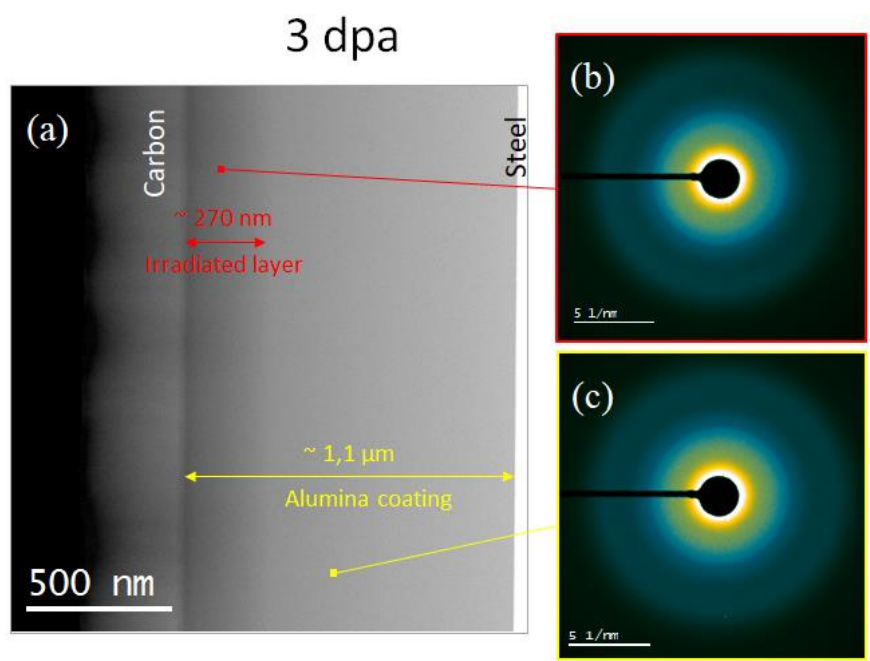

Fig. 8. (a) HAADF-STEM cross-sectional micrograph showing the alumina coating after 3 dpa ion irradiation (1.2 MeV Au$^+$) and corresponding DP's of: (b) irradiated layer and (c) unmodified coating volume.

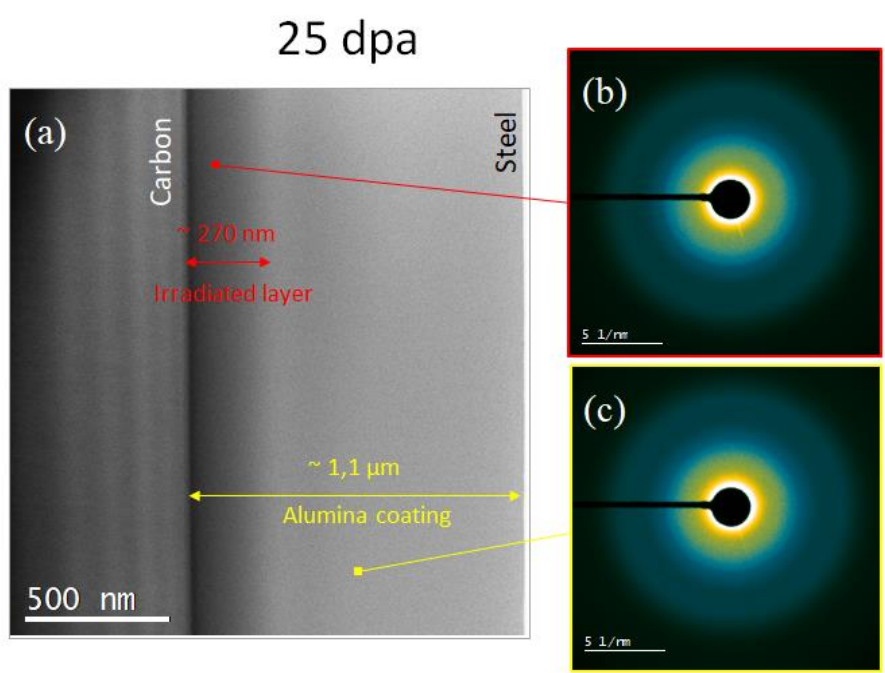

**Accepted Version**
Publication date: December 2021
Embargo: 24 Months
European Union, Horizon 2020, Grant Agreement number: 857470 — NOMATEN — H2020-WIDESPREAD-2018-2020
DOI: https://doi.org/10.1016/j.ceramint.2021.09.013

Fig. 9. (a) HAADF-STEM cross-sectional micrograph showing the alumina coating after 25 dpa ion irradiation (1.2 MeV $Au^+$) and corresponding DP's of: (b) irradiated layer and (c) unmodified coating volume.

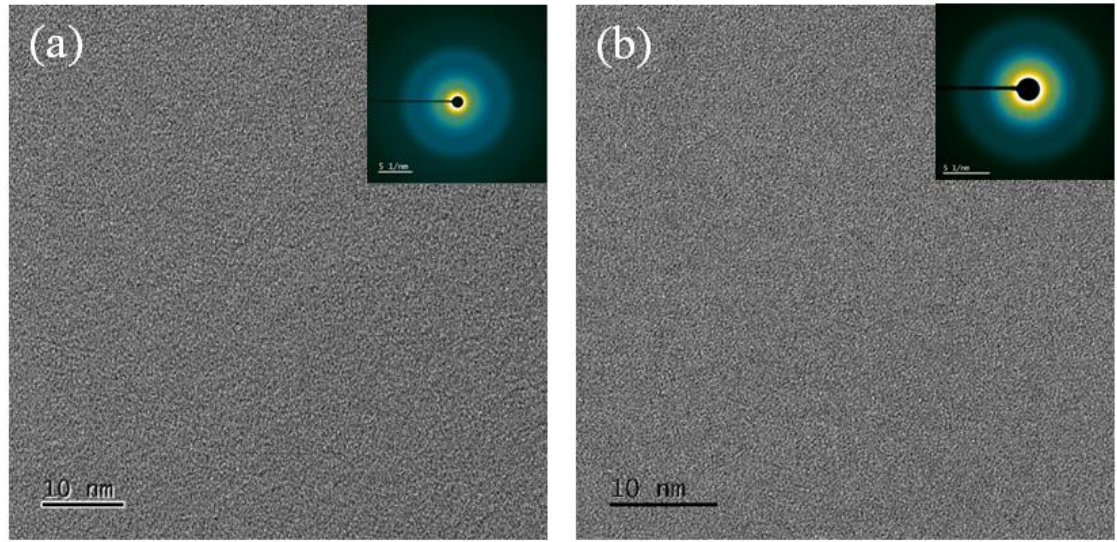

Fig. 10. HRTEM images and DP's insets showing the alumina coating surface: (a) before and (b) after 25 dpa ion irradiation (1.2 MeV $Au^+$). DPs were taken in the centers of the: (a) coating and (b) irradiated layer.

In order to better understand atomic ordering within irradiated zones, reduced distribution function (RDF) profiles were extracted from the collected SAED patterns. The software used for this purpose was eRDF Analyser [20]. According to our knowledge, such analysis is scarce in the literature and until now, no data have been published on ion irradiated alumina. An RDF analysis of electron diffraction intensities can provide information on average interatomic distances in materials [20] and is a useful tool to study the atomic short-range order in amorphous materials [20]. Recorded RDF profiles of implanted (red lines) and unimplanted (green lines) regions are shown in Fig.

**Accepted Version**
Publication date: December 2021
Embargo: 24 Months
European Union, Horizon 2020, Grant Agreement number: 857470 — NOMATEN — H2020-WIDESPREAD-2018-2020
DOI: https://doi.org/10.1016/j.ceramint.2021.09.013

11 (3 dpa sample) and Fig. 12 (25 dpa sample). Obtained data was normalized to the maximum peak intensity. The peaks at around 0.18 nm and 0.31 nm are indicative of Al-O and Al-Al bond lengths in amorphous $Al_2O_3$. These values are in good agreement with both the experimental results and the MD calculations reported in the literature [38–40]. Fig. 11 shows that for the 3 dpa sample (within measurement error), no significant differences between the RDFs of the implanted and unimplanted zones can be observed.

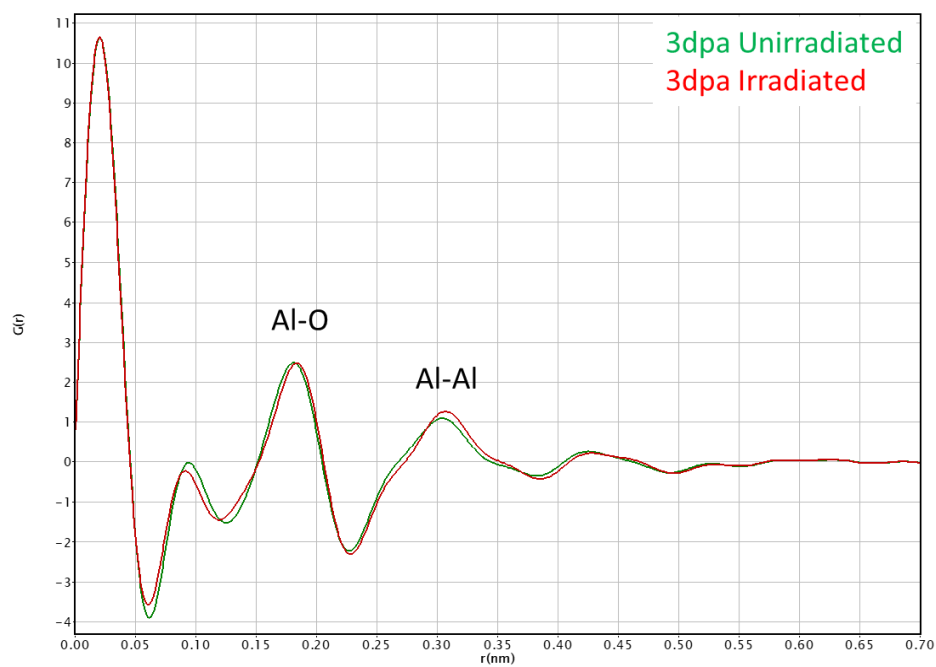

Fig. 11. The RDFs of the implanted (red line) and unimplanted (green line) zones of the 3 dpa alumina sample.

In the case of the 25 dpa sample (Fig. 12), some variations between the RDFs of the implanted and unimplanted zones were observed. For the implanted region, the relative intensity of Al-O and Al-Al bond peaks increases. In addition to that one can observe a slight relative shift in the Al-O peak position. Recorded data suggest that ion irradiation led to an increase of Al-O and Al-Al bond lengths. This phenomenon suggests

**Accepted Version**
Publication date: December 2021
Embargo: 24 Months
European Union, Horizon 2020, Grant Agreement number: 857470 — NOMATEN — H2020-WIDESPREAD-2018-2020
DOI: https://doi.org/10.1016/j.ceramint.2021.09.013

that local rearrangement of the atomic structure takes place. However, one should clearly explain that if this is true, observed occurrence concerns only atomic level range. This statement is supported by the shape of each bond peak (FWHM), which indicates that a long-range ordering does not take place. In conclusion, studied alumina coating should be regarded as amorphous material.

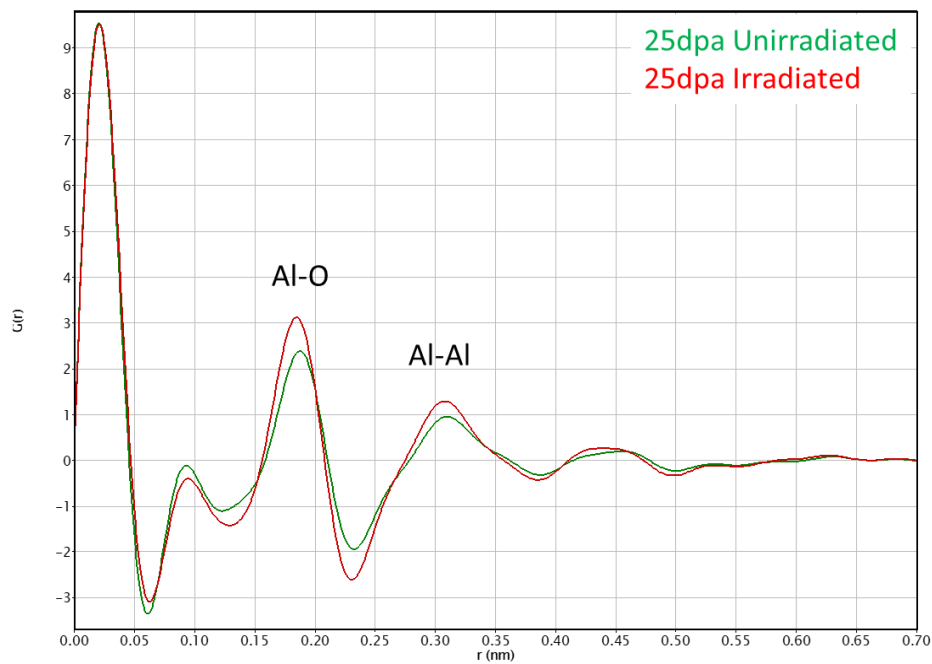

Fig. 12. The RDFs of the implanted (red line) and unimplanted (green line) zones of the 25 dpa alumina sample.

3.4. Grazing Incidence X-Ray diffraction (GIXRD)

GIXRD measurements were performed for virgin, 3 and 25 dpa samples. Fig. 13 shows recorded XRD patterns for virgin and ion irradiated material. In all tested specimens, only diffraction peaks characterized for steel substrate were observed. Obtained XRD clearly indicate that the alumina coatings remain amorphous before and

**Accepted Version**
Publication date: December 2021
Embargo: 24 Months
European Union, Horizon 2020, Grant Agreement number: 857470 — NOMATEN — H2020-WIDESPREAD-2018-2020
DOI: https://doi.org/10.1016/j.ceramint.2021.09.013

after ion irradiation (even for the highest damage of 25 dpa). However, one should remember, that according to our calculations, during XRD analysis, material is probed from about 2 um deep. Therefore, only about 50% of the signal originates from the alumina and 50% from the steel substrate. This is even less, in the case of ion irradiated layer. Finally, presented results support previously described TEM and RDF data, and points to the conclusion, that globally, the alumina coating remains amorphous even after submission to high damage level.  Lower intensity of the 110, 200, and 220 diffraction peaks, recorded for specimens irradiated up to 3 and 25 dpa, in comparison to the virgin specimen is most likely related to the variations in specimen flatness, positioning against the beam or smaller absorption in the non-irradiated layer.

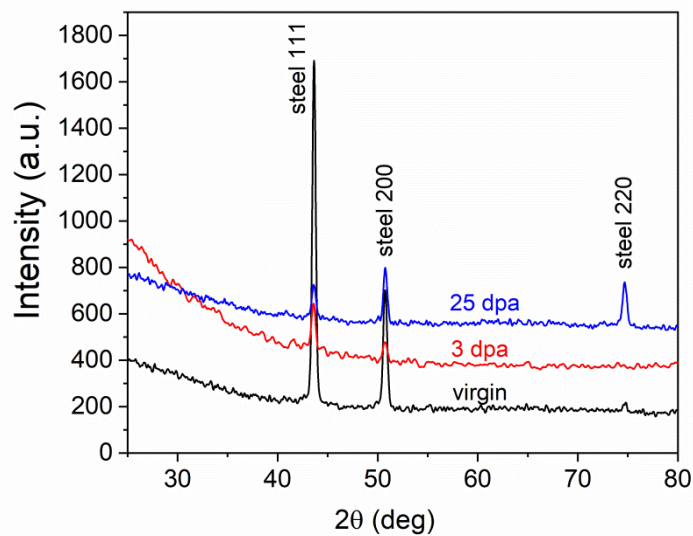

Fig. 13. X-ray diffraction of the virgin and ion irradiated up to 3 and 25 dpa alumina coating. Recorded diffraction peaks represent the steel substrate.

4. Discussion

**Accepted Version**

Publication date: December 2021
Embargo: 24 Months
European Union, Horizon 2020, Grant Agreement number: 857470 — NOMATEN — H2020-WIDESPREAD-2018-2020
DOI: https://doi.org/10.1016/j.ceramint.2021.09.013

This paper investigates the evolution of nanomechanical and structural properties of PLD-grown alumina coatings exposed to room temperature ion irradiation. Conducted nanoindentation tests revealed that the alumina-coated samples show some small reduction in hardness after ion irradiation. It was found that the hardness decrease is dose-dependent and reaches its peak at 3 dpa damage level. This is consistent with our previous findings, when ion irradiation was performed in the low energy range [19]. A comprehensive characterization combining TEM and GIXRD techniques provided insight into the nano- and micro- structure of the tested material. It was shown that the examined coating in the as-deposited state possesses an amorphous structure. No evidence of void formation, segregation of elements or crystallization were observed as a result of irradiation up to the highest dose of 25 dpa. These findings lead to the conclusion that under the given irradiation conditions, within the range of tested doses, the alumina coating remains amorphous, maintaining its structural integrity.

F. García Ferré and co-workers, in their groundbreaking report [11] investigated the evolution of the microstructure and mechanical properties of PLD-grown amorphous $Al_2O_3$ thin films subjected to ion irradiation at 600 °C. Published data showed that irradiation induces crystallization of the initially amorphous alumina phase. As crystalline alumina yields higher hardness (~25 GPa) than amorphous (~10-12 GPa) [41], hardness of the material increased as a result of phase transformation. In this study, no evidence of crystallization and the resulting hardening effect was observed. Since the lower irradiation temperature the higher dose threshold for crystallization is, it was expected

**Accepted Version**

Publication date: December 2021
Embargo: 24 Months
European Union, Horizon 2020, Grant Agreement number: 857470 — NOMATEN — H2020-WIDESPREAD-2018-2020
DOI: https://doi.org/10.1016/j.ceramint.2021.09.013

that at room temperature crystallization will not occur. The reported result confirms our predictions.

The decrease of the coating hardness after ion irradiation is not yet entirely understood. Several possible explanations can be given. First of all, we cannot rule out that increasing quantities of gold ions injected into the material with increasing irradiation dose might have influenced recorded mechanical property information. This is due to the fact that the indentation plastic zone extends deeper than the sample stopping-peak region [35,42,43]. Generally speaking, a very strong probability that registered hardness variation follows from chemical composition changes at the end of the implantation range exist. This conclusion is supported by the fact that irradiation was found to be free of the influence on Young's modulus. If this theory is correct, this would mean that for a given range of doses, ion irradiation has no effect on the hardness of the alumina coating. PDF analysis revealed that Al-Al and Al-O bond lengths increased slightly with Au irradiation. In fact, differences between registered RDFs could indicate that irradiated alumina is characterized by a higher degree of coordination. However, given that no significant variations in Young's modulus between tested samples were observed (or recorded changes are within measurement error), this hypothesis seems unlikely. On the other hand, the increase in bond distances could be attributed to the presence of the injected gold ions that push the target atoms away from each other. In order to confirm this hypothesis, currently we are performing advance simulation investigations to understand the impact of Au ions on the microstructure and its relation to mechanical properties.

**Accepted Version**
Publication date: December 2021
Embargo: 24 Months
European Union, Horizon 2020, Grant Agreement number: 857470 — NOMATEN — H2020-WIDESPREAD-2018-2020
DOI: https://doi.org/10.1016/j.ceramint.2021.09.013

Alternatively, the registered changes in alumina hardness could be interpreted as a result of the free volume content fluctuations. It is well known that in crystalline solids, the interactions between radiation and materials lead to the creation of the vacancy-interstitial pairs (so-called Frenkel pairs) which are central to radiation-induced effects [44]. The structure-property correlations of amorphous materials subjected to radiation are different from those of crystalline nature and are still not well understood. This knowledge gap is particularly pronounced for ceramic materials since most studies concerning the mechanical performance of disordered materials tend to focus on metallic glasses [45–50]. Unlike crystalline solids, amorphous materials lack long-range order and do not contain crystallographic point defects. Although they are microscopically isotropic and uniform, at the nanoscale their structure contains heterogeneities such as denser and looser atomic packing regions [51,52]. The loosely-packed fraction is termed "free volume" (FV) and has a lower atomic coordination than the reference region, which possesses a denser atomic packing. The low-temperature plastic deformation ability of amorphous solids is closely related to the FV content [51–54]. Some studies suggest that radiation can cause fluctuations in the FV concentration and thereby change the mechanical properties [45,48,49,55,56]. More specifically, it was shown that the increase in FV content results in a decrease of yield strength and hardness [45,55]. Obtained in this study hardness vs irradiation dose curve is in line with reported by Y. H. Qiu et al. [55] data. He studied the irradiation response of metallic glass $Ni_{50}Nb_{10}Zr_{15}Ti_{15}Pt_{7.5}Cu_{2.5}$ under 3 MeV $Au^+$ ion irradiation and linked registered hardness variations with fluctuations in free volume concentration. Nevertheless, the RDF findings of this study suggest that at the early stage of irradiation the short-range order in the material is nearly

**Accepted Version**

Publication date: December 2021
Embargo: 24 Months
European Union, Horizon 2020, Grant Agreement number: 857470 — NOMATEN — H2020-WIDESPREAD-2018-2020
DOI: https://doi.org/10.1016/j.ceramint.2021.09.013

unchanged. At the same time, for the same level of dpa, material experiences the greatest change in mechanical properties. In the light of above, there is no support for a hypothesis linking the registered reduction in hardness with the fluctuations of the free volume content. On the other hand, for the high dpa level, there are some subtle changes in the atomic-level structure of the material. It should be mentioned that our findings are based on a limited number of RDF data, and thus results from such analyses should be treated with considerable caution. Further work needs to be carried out to prove our postulates. Nevertheless, bearing in mind that the magnitude of the registered nanomechanical changes is not significant, it can be concluded that the coating is radiation-resistant in the evaluated experimental conditions.

Presented study complements previous research in the field of radiation resistance of amorphous materials [45,48,62,63,49,50,55,57–61]. One of the main issues in the knowledge in this field is the lack of well-established correlation between mechanical behavior and the atomic-level structure. This is due to the unique disordered and the nonequilibrium nature of amorphous solids, which is not entirely understood [64]. Irradiation effects may vary significantly between this group of materials. For example, in the literature, both radiation induced hardening caused by formation of nanocrystals [65] or changes in local bonding topology [66] and radiation induced softening associated with fluctuations in free volume content [48,55] were observed. A group of researchers from University of Nebraska revealed that other thin ceramic amorphous films, specifically SiOC retain glassy state under the range of irradiation temperatures and irradiation fluencies [57,59,66,67]. The group showed that ion irradiation does not cause

**Accepted Version**

Publication date: December 2021
Embargo: 24 Months
European Union, Horizon 2020, Grant Agreement number: 857470 — NOMATEN — H2020-WIDESPREAD-2018-2020
DOI: https://doi.org/10.1016/j.ceramint.2021.09.013

microstructure changes at large spatial scales. Interestingly, according to our knowledge, only scarce information's about the radiation damage resistance of amorphous coatings exist in the literature. Presented work enriches this area of research, points some pitfalls of the metal/oxide system and indicate potential research directions for the future.

5. Conclusion

This study has gone some way towards enhancing our understanding of the behavior of amorphous materials exposed to ion irradiation. The mechanical and structural response of thin PLD-grown alumina coatings on high energy (1.2 MeV $Au^+$) room temperature ion irradiation was investigated. Obtained nanomechanical results revealed that the hardness of the coating decreased slightly after implantation. Structural characterization shows that the coating possesses an amorphous structure and no evidence of crystallization was observed after irradiation (even up to 25 dpa). Potential mechanisms responsible for mechanical property changes were discussed. This work led to the conclusion that the coating exhibits excellent room temperature radiation tolerance, up to 25 dpa radiation damage level and presents promising evidence demonstrating the absolute radiation tolerance of amorphous PLD-grown alumina coatings.

Acknowledgments

The research leading to these results was carried out in the frame of the EERA Joint Programme on Nuclear Materials and is partly funded by the European Commission Horizon 2020 Framework Programme under grant agreement No. 755269 (GEMMA project). Also, financial support from the National Centre for Research and Development

**Accepted Version**

Publication date: December 2021
Embargo: 24 Months
European Union, Horizon 2020, Grant Agreement number: 857470 — NOMATEN — H2020-WIDESPREAD-2018-2020
DOI: https://doi.org/10.1016/j.ceramint.2021.09.013

through a research grant "Studies of the role of interfaces in multi-layered, coated and composite structures" PL-RPA2/01/INLAS/2019 is gratefully acknowledged. The financial support of the National Research Foundation and Department of Science and Innovation in South Africa is gratefully acknowledged. The Research Council of Norway is acknowledged for the support to the Norwegian Micro- and Nano-Fabrication Facility, NorFab, project number 295864.

**Accepted Version**

Publication date: December 2021
Embargo: 24 Months
European Union, Horizon 2020, Grant Agreement number: 857470 — NOMATEN — H2020-WIDESPREAD-2018-2020
DOI: https://doi.org/10.1016/j.ceramint.2021.09.013

**Accepted Version**

Publication date: December 2021
Embargo: 24 Months
European Union, Horizon 2020, Grant Agreement number: 857470 — NOMATEN — H2020-WIDESPREAD-2018-2020
DOI: https://doi.org/10.1016/j.ceramint.2021.09.013

**Accepted Version**

Publication date: December 2021
Embargo: 24 Months
European Union, Horizon 2020, Grant Agreement number: 857470 — NOMATEN — H2020-WIDESPREAD-2018-2020
DOI: https://doi.org/10.1016/j.ceramint.2021.09.013

**Accepted Version**

Publication date: December 2021
Embargo: 24 Months
European Union, Horizon 2020, Grant Agreement number: 857470 — NOMATEN — H2020-WIDESPREAD-2018-2020
DOI: https://doi.org/10.1016/j.ceramint.2021.09.013

**Accepted Version**

Publication date: December 2021
Embargo: 24 Months
European Union, Horizon 2020, Grant Agreement number: 857470 — NOMATEN — H2020-WIDESPREAD-2018-2020
DOI: https://doi.org/10.1016/j.ceramint.2021.09.013